\documentclass{PoS}

\usepackage{graphicx}
\PoS{PoS(LAT2005)166}

\title{The 3-state Potts model as a heavy quark finite density laboratory }

\ShortTitle{The 3-states Potts model}

\author{\speaker{Seyong Kim}\thanks{S.K. is supported by Grant
        No. R01-2002-000-00291-0 from the Basic Research Program of
        Korea Science and Engineering Foundation}\\ 
        Department of Physics, Sejong University, Seoul 143-747,
        Korea\\
	Department of Physics, U. of Wales Swansea, Swansea SA2
        8PP, U.K. \\ 
	E-mail: \email{skim@sejong.ac.kr}} 

\author{Ph. de Forcrand \\
        Institut f\"ur Theoretische Physik, ETH-Z\"urich,
        CH-8093 Z\"urich, Switzerland\\
        CERN Theory Division, CH-1211 Geneva 23, Switzerland\\
        E-mail: \email{forcrand@phys.ethz.ch}}

\author{S. Kratochvila \\
        Institut f\"ur Theoretische Physik, ETH-Z\"urich,
        CH-8093 Z\"urich, Switzerland\\
        E-mail: \email{skratoch@phys.ethz.ch}\\
        }

\author{T. Takaishi \\
        Hiroshima U. of Economics \\
        Hiroshima 731-0124, Japan\\
        E-mail: \email{takaishi@hiroshima-u.ac.jp} \\
        }

\abstract{The 3-D Z(3) Potts model is a model for finite temperature QCD
  with heavy quarks. The chemical potential in QCD becomes an external
  magnetic field in the Potts model. Following Alford et
  al.\cite{Alford_et_al}, 
  we revisit this mapping, and determine the phase diagram for an
  arbitrary chemical potential, real or imaginary. Analytic
  continuation of the phase transition line between real and imaginary
  chemical potential can be tested with precision. 
  Our results show that the chemical potential weakens the heavy-quark
  deconfinement transition in QCD.
}

\FullConference{XXIIIrd International Symposium on Lattice Field Theory\\
25-30 July 2005\\
Trinity College, Dublin, Ireland}

\begin{document}

\section{Introduction}

The 3-dimensional 3-state Potts model and the finite temperature 
QCD with infinitely heavy quarks share the same $Z(3)$ global 
symmetry, and in both theories the order parameter shows a first 
order phase transition. When an external magnetic field is turned 
on in the Potts model, this $Z(3)$ symmetry is explicitly broken 
and the first order phase transition is weakened, becoming second order 
for a certain strength of the magnetic field and crossover beyond. In the 
case of finite temperature QCD, the quark mass plays the role of 
the external magnetic field in the 3-D 3-state Potts model. The first order 
phase transition of the quenched theory (infinitely heavy quark limit) weakens,
and turns into a second order transition for a critical, heavy but not
infinite, dynamical quark mass \cite{Hasenfratz, Alexandrou_et_al}. A
universality argument then implies that the critical properties of the
two theories, Potts and QCD, will be the same at this second-order
point. Since the 3-D 3-state Potts model is simpler than QCD and can
be simulated on a large lattice, its numerical investigation will give
us more detailed quantitative information on the critical properties
of finite temperature QCD with heavy quarks.

The effect of a quark chemical potential on hot QCD with static quark
sources can be formulated in the following grand canonical ensemble
\cite{Alford_et_al}: 
\begin{eqnarray}
Z(\mu) &=& \sum_{n,\overline{n}} Z_{n,\overline{n}} \;\; e^{\beta \mu (n -
  \overline{n})} \nonumber \\
&=& \sum_{n,\overline{n}} \int [{\cal D} U] \;\; \frac{1}{n!}
  (\Phi [U])^n \frac{1}{\overline{n}!} \;\; (\Phi [U]^*)^{\overline{n}} \;\;
  e^{-S_g[U] -  \beta n (M - \mu) - \beta \overline{n} (M + \mu)} \nonumber \\
&=& \int [{\cal D} U] \exp ({-S_g [U] + e^{-\beta (M-\mu)} \Phi [U] +
  e^{-\beta (M+\mu)} \Phi[U]^*}), \label{grand}
\end{eqnarray}
where $Z_{n,\overline{n}}$ is the canonical partition function with
$n$ quarks and $\overline{n}$ anti-quarks, $U$ is the SU(3) gauge field,
$S_g$ is the gauge action, $\Phi$ is the Polyakov line, $\Phi^*$ the
anti-Polyakov line, $M$ is the heavy quark mass and $\mu$ is the quark
chemical potential. When there is no chemical potential ($\mu=0$), the
action in (\ref{grand}) is real. If $\mu \neq 0$, the action becomes
complex and Monte Carlo simulation is difficult because of the ``sign
problem'': the usual probabilistic interpretation of Eq. (\ref{grand})
is not possible. 

Symmetry consideration tells us that the important dynamics of the
gauge field are those of the Polyakov line and the anti-Polyakov line.
Thus, the corresponding lattice Hamiltonian for the 3-D 3-state Potts
model is given by: 
\begin{equation}
H = - k \sum_{i,\vec{x}} \delta_{\Phi
    (\vec{x}), \Phi (\vec{x}+i)} - \sum_{\vec{x}} [ h
  \Phi (\vec{x}) +   h' \Phi^* (\vec{x})]
\label{ZPotts}
\end{equation}
where $h=e^{-\beta(M - \mu)} = h_M e^{\beta \mu} = h_M
e^{\overline{\mu}}$ and $h' = e^{-\beta(M + \mu)} = h_M e^{-\beta
\mu}= h_M e^{-\overline{\mu}}$ with $h_M = e^{-M/T}$,
$\overline{\mu}=\mu/T$, and $\Phi (\vec{x})$ is a $Z(3)$ spin. If $h
\neq {h'}^*$, this Hamiltonian also is complex. However, the partition
function remains real \cite{Alford_et_al}. The partition function
includes summation over all the possible $Z(3)$ spin
configurations. Introducing ``bonds'' with a certain probability among
parallel $Z(3)$ spins and defining a ``cluster'' made of the sites
connected by the bonds, we can divide the summation into the sum over
clusters and that within a cluster. Then, after analytically summing
over $Z(3)$ spin orientations within each cluster, the partition
function in terms of cluster configurations is given by
\begin{eqnarray}
Z = \int [{\cal D} b] (e^k - 1)^{N_b} \prod_{C} \left[e^{2 h_M |C|
    \cosh\overline{\mu}} + 2 e^{- h_M |C| \cosh \overline{\mu}}
    \cos (\sqrt{3} h_M |C| \sinh \overline{\mu}) \right],  \label{Zbond}
\end{eqnarray}
where ${\cal D} b$ is the sum over all the possible bond
configurations, $N_b$ is the number of bonds in a given cluster
configuration, and $|C|$ is the number of sites in a given
cluster. $Z$ is real and is free from the ``sign problem'' since the
second term which can be negative due to the presence of $\cos
(\sqrt{3} h_M |C| \sinh \overline{\mu})$ is always smaller than the
first term. However, the ``solution'' of the complex action problem in
the 3-D 3-state Potts model is different from the $SU(2)$ gauge theory
case \cite{Pietro} and from the four-fermi theory case \cite{Azcoiti},
where the action itself can be shown to be real even with a real
chemical potential. In the Potts model, the action is complex, and a
change of variables (from spins to bonds) is necessary to recover a
real effective action and show that the partition function of the
model remains real.

The ($h = h'$) case (that is, zero chemical potential)
has been studied in \cite{Karsch_Stickan}. The ($h' = 0$) case (that is,
$M, \mu \rightarrow \infty $ while $M/\mu$ is fixed) has been
investigated in \cite{Alford_et_al}. Since we are also interested in testing
the analytic continuation of imaginary chemical potential results to real
chemical potential, instead of taking $h' = 0$ limit, we study here
the full parameter space of arbitrary $(h,h')$.

For the case of an imaginary chemical potential ($\mu
\rightarrow \mu_I$, i.e., $ h' = h_M e^{-i\overline{\mu}_I} = h^*$),
the partition function, $Z_I$, becomes 
\begin{eqnarray}
Z_I = \int {\cal D} b (e^k - 1)^{N_b} \prod_{C} \left[e^{2 h_M |C|
    \cos\overline{\mu}_I} + 2 e^{- h_M |C| \cos \overline{\mu}_I}
    \cosh (\sqrt{3} h_M |C| \sin \overline{\mu}_I) \right] .  \label{ZIbond}
\end{eqnarray}
Since $\cosh (i \overline{\mu}) = \cos (\overline{\mu}), \sinh (i
\overline{\mu}) = i \sin (\overline{\mu})$ and $\cos (i \theta) =
\cosh (\theta)$, the relation $Z (\mu\rightarrow i \mu_I ) = Z_I
(\mu_I)$ is quite obvious and the analytic continuation between $Z$
and $Z_I$ is transparent. The Roberge-Weiss symmetry \cite{Roberge_Weiss},
$Z_I(\frac{\mu_I}{T}) = Z_I(\frac{\mu_I}{T} + \frac{2\pi}{3} n)$, 
can also be seen clearly in (\ref{ZIbond}). 

\section{Real Chemical Potential}

Using the Swendsen-Wang cluster algorithm \cite{S_W} on (\ref{Zbond}),
we simulate the 3-D 3-state Potts model. The actual simulation is
performed along the $h = h'$ line in the $(h,h')$ parameter space and
the data are re-weighted for arbitrary $(h,h')$. Although the
reweighting factor is not always positive, this strategy is more
efficient than sampling Eq.~(\ref{Zbond}), because the ensuing sign
problem is very mild. Simulations are performed on lattice volumes
$56^3, 64^3$, and $72^3$. Typically $\sim$ 2 million data sample is
collected. The critical point is located by requiring the third order
Binder cumulant of the $Z(3)$ spin magnetization to vanish and by
evaluating the fourth order Binder cumulant ($B_4$) of the
magnetization at that point. Since the universality class of the 3-D
3-state Potts model is that of the 3-D Ising model, $B_4$ for the
Potts model at the critical point should be equal to that of the Ising
model ($= 1.604(1)$). Figure (\ref{B4}) shows the fourth order Binder
cumulant as a function of $h$ (here, $h' = h$). From this Figure, we
find that the critical end point is at $(k_c,h_c) = (0.54940(4)
0.000255(5))$ in comparison with the value given in
\cite{Karsch_Stickan}, $(0.54938(2),0.000258(3))$. Similarly, for the
$h' = 0$ case, we obtain that $(k_c,h_c) = (0.54947(1),0.000465(5))$
(\cite{Alford_et_al} reported $(k_c,h_c) =
(0.549463(13),0.000470(2))$.

\begin{figure}[ht]
\vskip 0.5cm
\begin{center}
\includegraphics[width=2.8in]{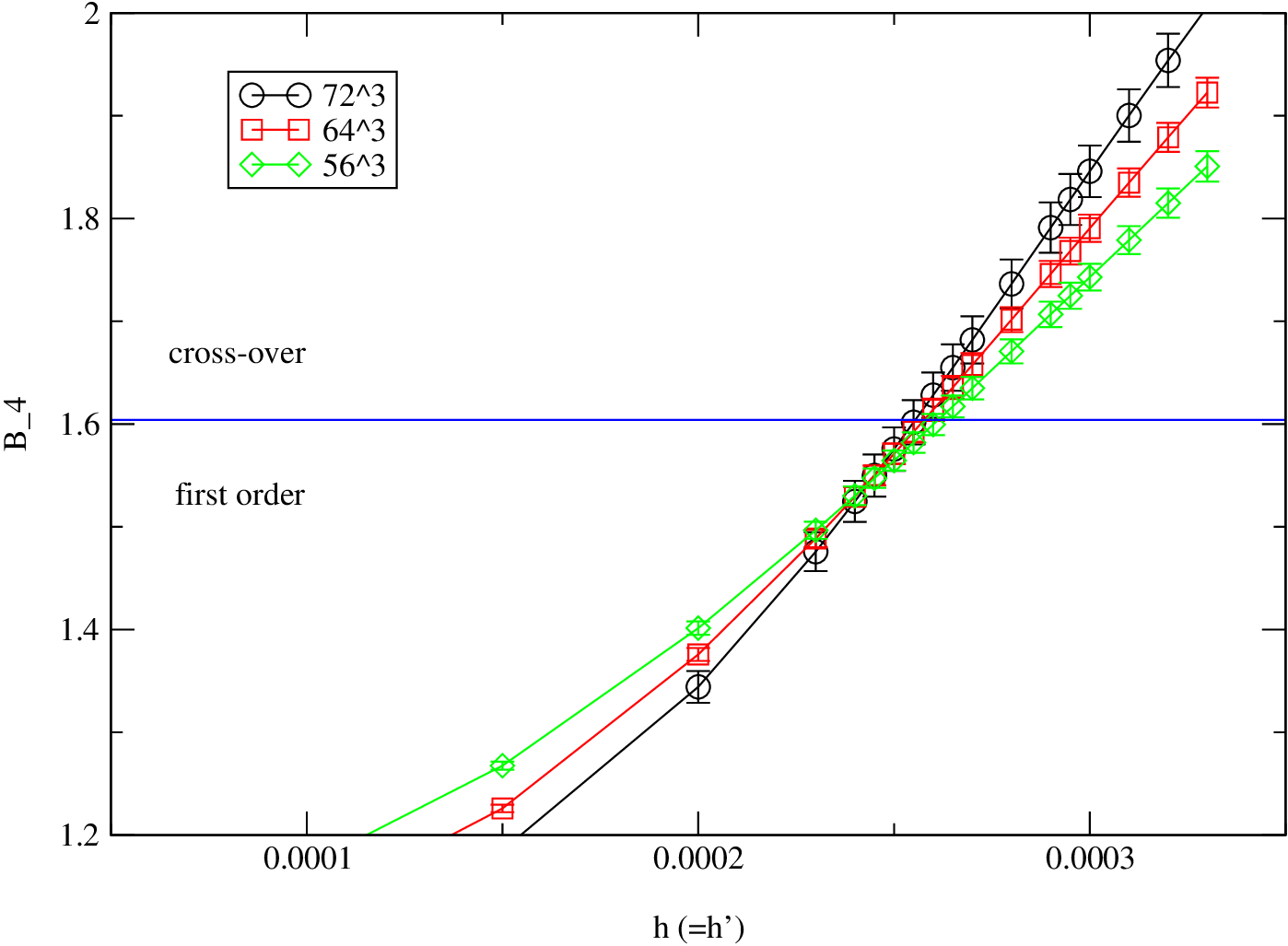}
\hskip 0.8cm
\includegraphics[width=2.8in]{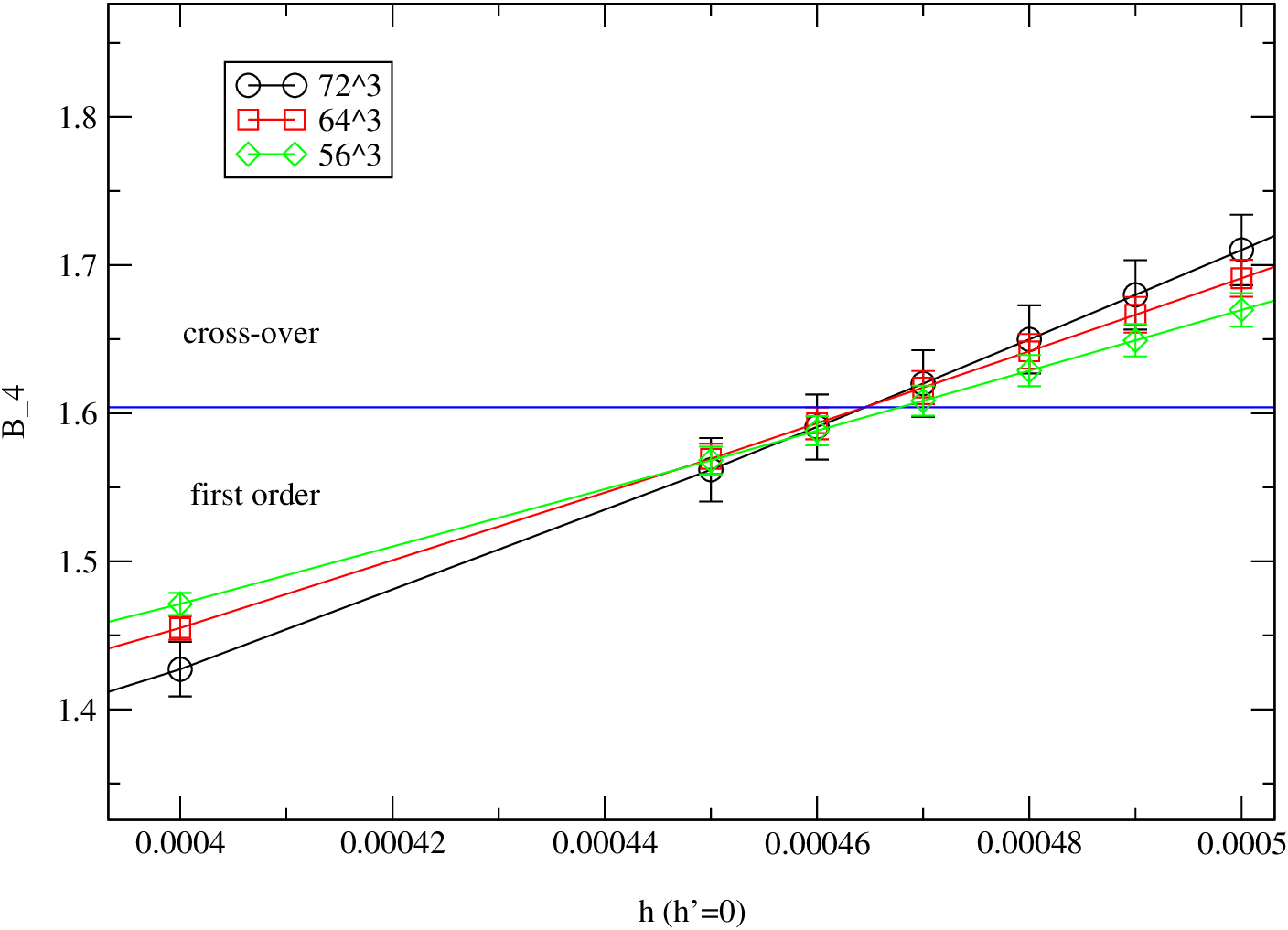}
\caption{$B_4$ for magnetization vs $h$ ($h=h'$) \hskip
  1.5cm {\bf Figure 2:} $B_4$ for magnetization vs $h$ ($h'=0$)}
\label{B4}
\end{center}
\setcounter{figure}{2}
\end{figure}

Similarly, the critical end point can be located for arbitrary $h$ and
$h'$. Figure (\ref{kc}) shows the $(h,h')$ parameters for the second
order phase transition. Here, instead of directly plotting $(h,h')$, 
we use variables $M/T$ and $\mu/T$, which can be related to QCD. 
The line in the figure is from the asymptotic value, $(h =
h_c,h'= 0)$. Since $h = e^{-M/T + \mu/T}$, the relation $h_c = e^{-M/T
+ \mu/T}$ for given $h_c$ defines a line in the parameter space
$(\frac{M}{T}, \frac{\mu}{T})$. In Figure (3), the simulation data
lies above this asymptotic line. This means that the anti-Polyakov line
in finite temperature QCD plays an important role in critical
parameter space and should not be neglected in considering the
critical properties of the theory. To further compare our results with the
limit studied in \cite{Alford_et_al}, we show $T/M$ vs. $\mu/M$ in
Figure (4). As $\mu/M$ increases, our simulation data approach the
asymptotic line as expected. This figure clearly shows that the range
of $T/M$ values for which the transition is first-order shrinks as the
chemical potential is turned on. In fact, the transition disappears
altogether for $\frac{\mu}{M} > 1$. So the effect of the chemical
potential is to {\em weaken} the phase transition.

\begin{figure}[ht]
\begin{center}
\includegraphics[width=2.8in]{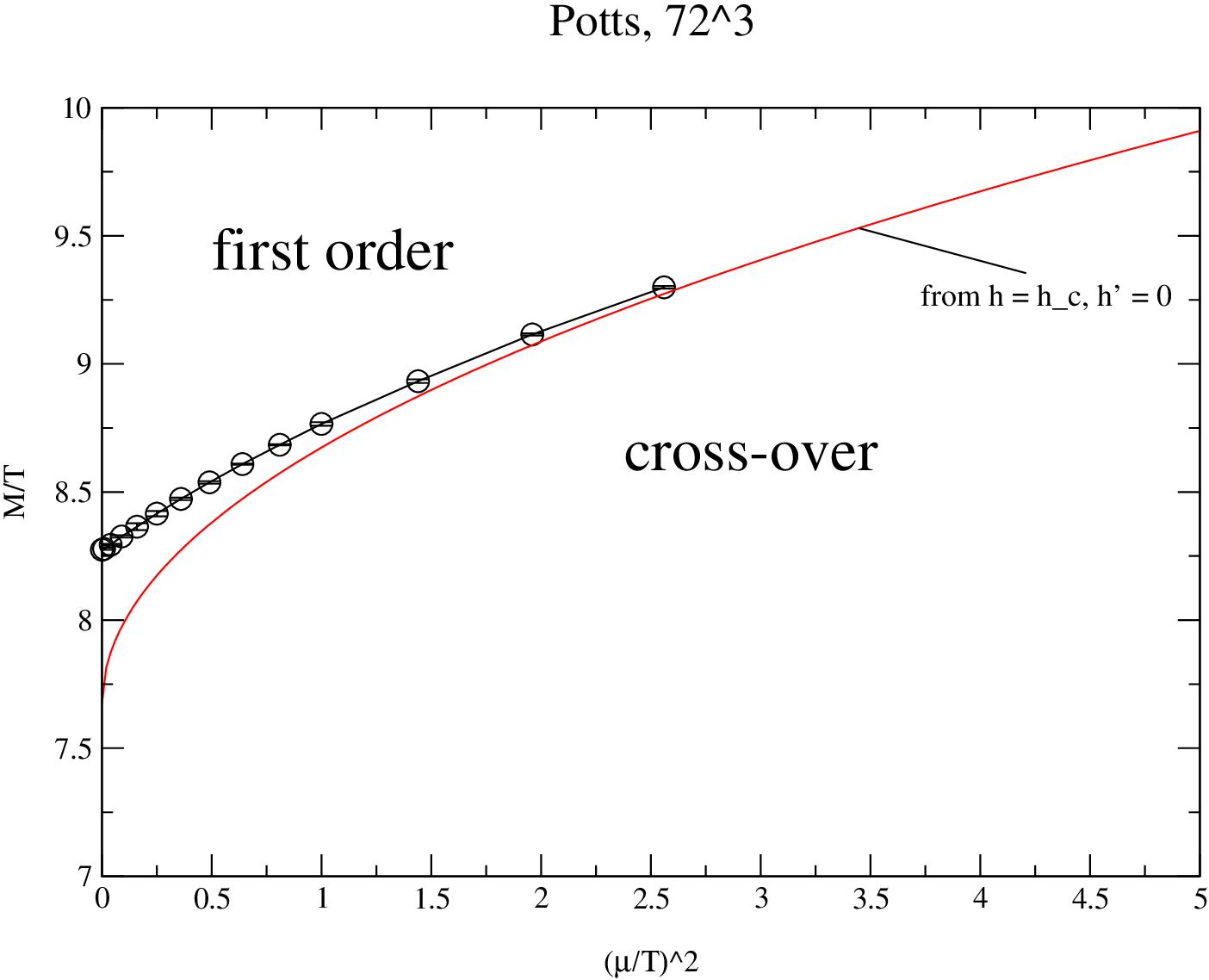}
\hskip 0.5cm
\includegraphics[width=2.8in]{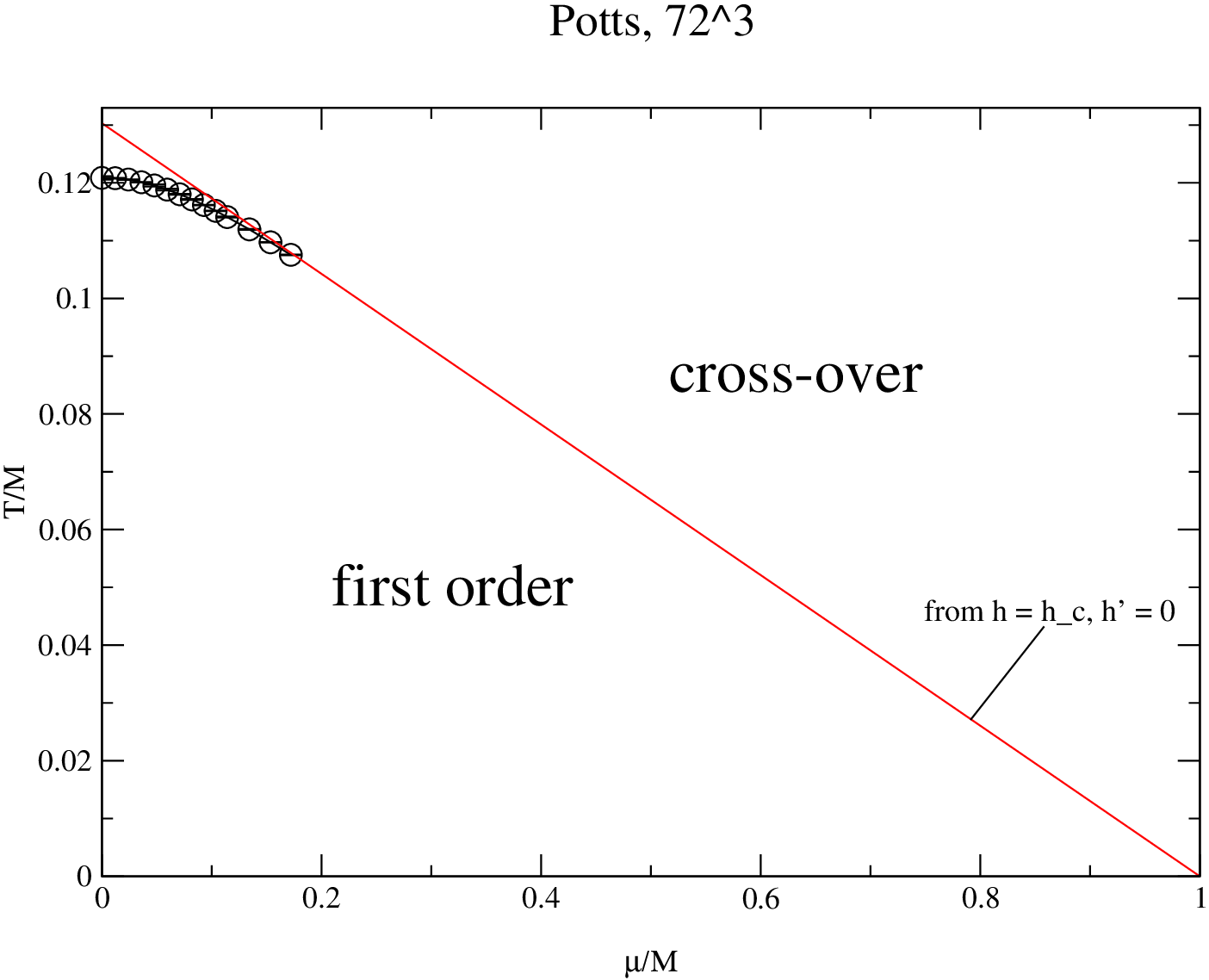}
\vskip 0.3cm
\caption{$M/T$ for 2nd order transition vs. $(\mu/T)^2$\hskip 1.0cm
  {\bf Figure 4:} $T/M$ for 2nd order transition vs. $\mu/M$}
\label{kc}
\end{center}
\vskip 0.3cm
\setcounter{figure}{4}
\end{figure}

Figure (\ref{kc}) suggests an interesting comparison with the finite
temperature QCD phase diagram. Let us imagine increasing the chemical
potential from zero, following the arrow in Figure (5). Without chemical
potential, the quark mass is chosen such that the transition is first order. 
With a non-zero chemical potential, this first order phase
transition weakens. At a certain chemical potential, the system
shows a second order phase transition. With chemical potential larger
than this critical value, the system shows a cross-over. On the other
hand, consider Figure (6) which is analogous to Figure (5). This
figure is a schematic phase diagram for three-flavor QCD with {\em
  light} quarks suggested in \cite{Owe_Philippe} and summarizes the
conventional wisdom. Without chemical potential, 3-flavor QCD shows a
first order phase transition when the quark mass is smaller than a
certain mass ($m_c (0)$). Let us say that we start with a system where
the quark mass is larger than $m_c (0)$. Then, this 3-flavor QCD
system has a cross-over. In this case, increasing the chemical
potential makes the transition stronger: a second order phase
transition appears when the chemical potential hits a certain
magnitude. If the chemical potential is increased further, the system
undergoes a first order phase transition. Therefore, the 3 light
flavor QCD system shows (cross-over $\rightarrow$ second order phase
transition $\rightarrow$ first order phase transition) as the chemical
potential is increased. In contrast, the heavy quark QCD system
suggested by 3-D $Z(3)$ Potts model will undergo (first order phase
transition $\rightarrow$ second order phase transition $\rightarrow$
cross-over) as the chemical potential is increased. Interestingly,
some recent light-quark QCD simulations give support to the
possibility that the conventional wisdom scenario above might be at
fault \cite{Owe_plenary}, so that the influence of the chemical
potential in the light-quark and the heavy-quark cases would be
similar.

\begin{figure}[ht]
\begin{center}
\includegraphics[width=2.8in]{musqvsmdt5.eps}
\hskip 1.0cm
\includegraphics[width=2.6in]{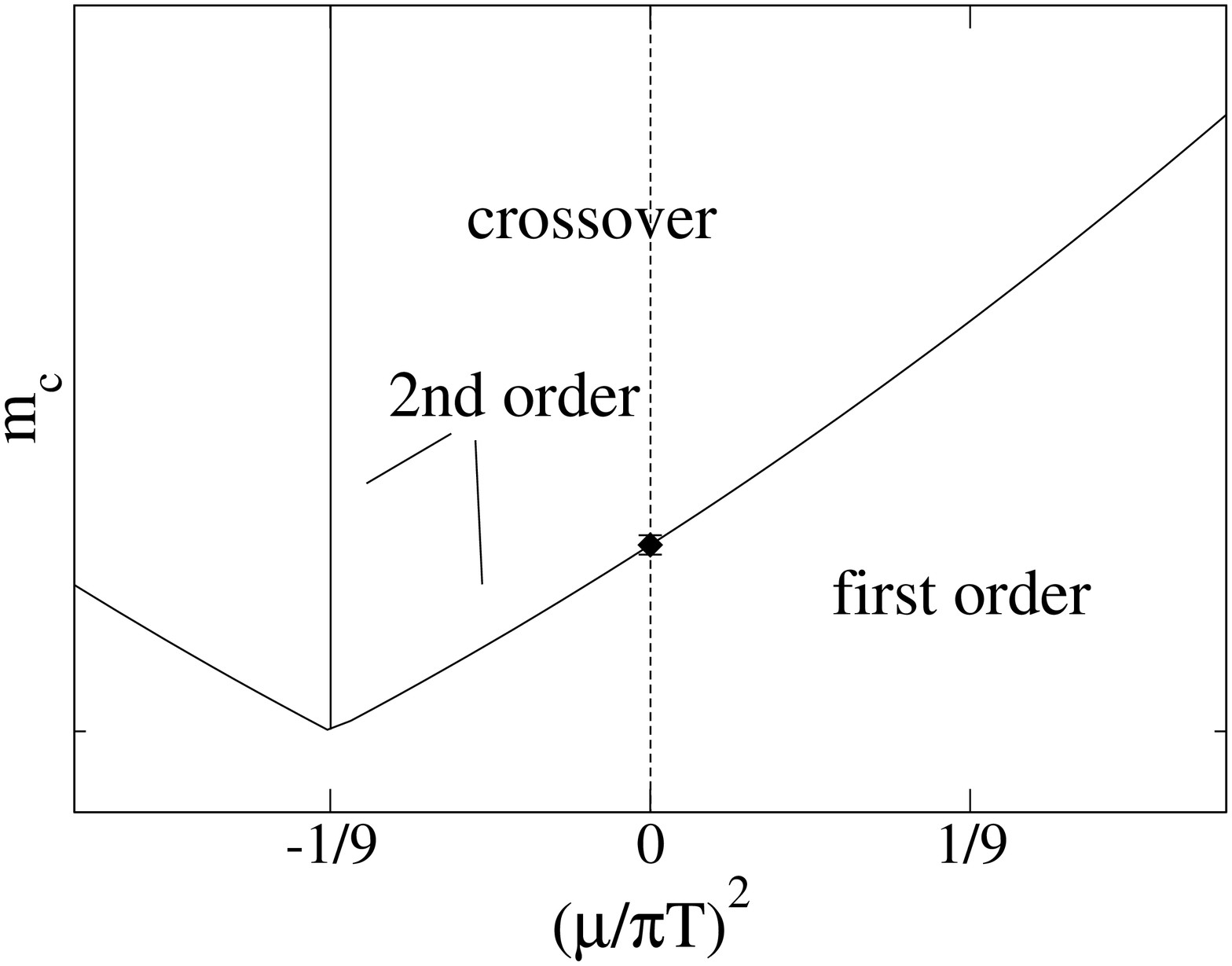}
\end{center}
\caption{$M/T$ for critical point vs. $(\frac{\mu}{T})^2$ \hskip 0.5cm
    {\bf Figure 6:} schematic QCD phase diagram for 3 light quarks}
\setcounter{figure}{4}
\label{im}
\end{figure}

\section{Imaginary vs Real Chemical Potential}

For the case of an imaginary chemical potential, one can perform a
direct sampling of Eq.~(\ref{ZPotts}), or reweight the $h = h'$ data
used for the real chemical potential study. As before, the critical
point is located by use of the magnetization Binder cumulant and $B_4
= 1.604$ crossing point.

In Figure (\ref{im}), we put together the critical end point
parameters obtained from the imaginary chemical potential case and
those from the real chemical potential case (note the similarity
with the schematic phase diagram for 3 light flavor QCD
shown in Figure (6)). Although the proximity of the Roberge-Weiss
transition at $\mu_I = \pi/3$ introduces curvature in the imaginary
chemical potential result, the small real $\mu$ result is smoothly
connected to the small imaginary $\mu_I$ result and the analytic
continuation poses no serious problem even on a large lattice volume
such as $72^3$. However, the critical line shows significant curvature,
which limits the accuracy of the extrapolation from imaginary to real
$\mu$. A 4th order polynomial $\frac{M}{T} = 8.273 + 0.585
(\frac{\mu}{T})^2 - 0.174 (\frac{\mu}{T})^4 + 0.160 (\frac{\mu}{T})^6
- 0.071 (\frac{\mu}{T})^8$ is necessary to describe the critical line
from $(\frac{\mu}{T})^2 = -(\frac{\pi}{3})^2$ to 1.5. 

\section{Summary}

We extend earlier work on the 3-D $Z(3)$ Potts model with one external
field coupled to $\Phi (\vec{x}) + \Phi^* (\vec{x})$
\cite{Karsch_Stickan} or to $\Phi (\vec{x})$ \cite{Alford_et_al} to
the full general case $h \Phi (\vec{x}) + h' \Phi^* (\vec{x})$, and
present a complete picture of the phase diagram in the $(\kappa,h,h')$
parameter space.  Our investigation also gives us a handle on heavy
quark QCD at finite density and temperature which shares the same
global symmetry, allowing us to make statements about the heavy quark
QCD phase diagram in the finite temperature, finite density, and heavy
quark mass parameter space.

For a real chemical potential, the partition function of the Potts
model is shown to be real even though the action itself is complex.
The sign problem is mild, and simulation results show that turning on
a chemical potential $\mu$ makes the finite temperature transition
weaker, so that the region of parameter space corresponding to a
first-order transition shrinks under the influence of $\mu$. This
implies that a similar phenomenon occurs in the phase diagram of heavy
quark QCD.

For an imaginary chemical potential, the action of the Potts model is
real and the model is easy to study. Since both real chemical
potential and imaginary chemical potential can be simulated, analytic
continuation from imaginary to real chemical potential can be tested.
We find that analytic continuation works satisfactorily, even with
large lattice volumes such as $72^3$.

In short, the 3-D $Z(3)$ Potts model has a rich structure and provides us
with a useful ``proving ground'' for studying the finite temperature/density
phase structure of QCD.

\end{document}